\documentclass[aps,prl,twocolumn,superscriptaddress]{revtex4}
\usepackage{graphicx}

\begin{document}

\title{Exchange interaction effects on the optical properties of LuMnO$_3$}

\author{A. B. Souchkov}
\author{J. R. Simpson}
\affiliation{Materials Research Science and Engineering Center, 
University of Maryland, College Park, Maryland 20742, U.S.A.}
\author{M. Quijada}
\affiliation{NASA, Goddard Space Flight Ctr, Code 551, Greenbelt, 
Maryland 20771, U.S.A.}
\author{H. Ishibashi\footnote{Present address: Dept. of Materials Science, Division of Science,
Osaka Prefecture University 1-1 Gakuen-cho, Sakai, Osaka 599-8531, Japan.}} 
\affiliation{Department of Physics and Astronomy, Rutgers 
University, Piscataway, New Jersey 08854, U.S.A.}
\author{N.~Hur}
\affiliation{Department of Physics and Astronomy, Rutgers University, 
Piscataway, New Jersey 08854, U.S.A.}
\author{J.~S. Ahn}
\affiliation{Department of Physics and Astronomy, Rutgers University, 
Piscataway, New Jersey 08854, U.S.A.} \affiliation{Center for Strongly 
Correlated Materials Research, Seoul National University, Seoul 
151-742, Republic of Korea}
\author{S. W. Cheong}
 \affiliation{Department of Physics and Astronomy, Rutgers University, Piscataway, New Jersey 08854, U.S.A.}
\author{A. J. Millis}
\affiliation{Department of Physics, Columbia University, 538 W 120 St, 
New York, New York 10027, U.S.A.}
\author{H. D. Drew}
\affiliation{Materials Research Science and Engineering Center, 
University of Maryland, College Park, Maryland 20742, U.S.A.}


\date{\today}

\begin{abstract}
We have measured the optical conductivity of single crystal LuMnO$_3$ 
from 10~to~45000~cm$^{-1}$ at temperatures between 4 and 300~K. A 
symmetry allowed on-site Mn $d$-$d$ transition near 1.7~eV is observed 
to blue shift ($\sim$0.1~eV) in the antiferromagnetic state due to 
Mn--Mn superexchange interactions.  Similar anomalies are observed in 
the temperature dependence of the TO phonon frequencies which arise 
from spin-phonon interaction. We find that the known anomaly in 
temperature dependence of the quasi-static dielectric constant 
$\epsilon_0$ below the $T_N \sim 90$~K is overwhelmingly dominated by  
the phonon contributions.

 
\end{abstract}


\maketitle

The colossal magnetoresistance compounds based on doped pseudo-cubic 
LaMnO$_3$ have excited much attention because of their interesting 
physical properties and potential applications. Another series of 
$R$MnO$_3$ materials ($R$=Ho,Er,Tm,Yb,Lu, or Y,Sc,In) have smaller 
radius $R^{3+}$ ions and crystallize in the hexagonal lattice. The 
hexagonal manganites are interesting as examples of multiferroics (or 
ferroelectromagnets) \cite{Smolenskii1} —-- they are both ferroelectric 
($T_c$$\sim$900~K) and strongly frustrated antiferromagnets 
($T_N$$\sim$90~K). The coupling between ferroelectric and magnetic 
order parameters provides the prospect of manipulating electrical 
properties through magnetic fields and vice versa which, in turn, 
gives these compounds potential for applications in electronics.

There have been several reports on the magnetic structure 
\cite{Bertaut2,Munoz1,Fiebig1} and aspects of the electromagnetic 
response in hexa-manganites 
\cite{Fiebig1,Kritayakirana1,Penney1,Iliev1,Yi1,Takahashi1}. Of 
particular interest is the observation of a temperature anomaly of the 
static dielectric constant, $\epsilon_0$, below $T_N$.  In this 
letter, we present a systematic study of the linear optical response 
of LuMnO$_3$ which elucidates the origin of the $\epsilon_0$ anomalies 
and the effects of the Mn--Mn exchange energy on the electrodynamics 
of this system.  We show that the exchange interaction manifests 
itself in an antiferromagnetic resonance, spin-phonon coupling, and 
the temperature dependence of a 1.7~eV on-site Mn $d$-$d$ optical 
transition.  The results give a comprehensive view of the magnetic and 
electronic structure of this interesting ferroelectric and strongly 
frustrated antiferromagnetic material.

The high-temperature paraelectric phase in the hexagonal manganites 
may be considered as layers of corner-sharing MnO$_5$ triangular 
bipyramids connected by a layer of $R$-ions. In the ferroelectric 
phase, $R$-ions alternate their $c$-axis coordinates, producing a net 
electric moment of the unit cell, and the MnO$_5$ pyramids tilt away 
from the $c$-axis. The $d$-orbitals of the Mn$^{3+}$ ion in a 
triangular bipyramid of five O$^{2-}$ ions are split by the crystal 
field into three groups: $d_{xz,yz}, d_{xy,x^2-y^2}$, and 
$d_{3z^2-r^2}$, in order of increasing energy. The latter orbital has 
the highest energy as the Mn-apical O bond lengths are shorter than 
in-plane distances. The in-plane $d_{xy,x^2-y^2}$ orbitals are 
strongly hybridized with oxygen $p$ orbitals. In the ground state of 
the Mn$^{3+}$ ion four electrons occupy four lowest orbitals giving 
$<L>=0$ and $S=2$. Below $T_N$, spins are ordered 
antiferromagnetically in-plane \cite{Munoz1,Fiebig1}. The Neel 
temperature in hexa-manganites is 6--10 times smaller than the Weiss 
temperature which is due to the spin frustration in the triangular 
planar lattice and weak inter-plane exchange interaction.

Single crystals of LuMnO$_3$ were grown using the travelling floating 
zone method and characterized by magnetization, resistivity, and x-ray 
powder diffraction. The lattice constants as well as the observed 
macroscopic properties agree well with measurements reported in 
literature. Platelet samples were cleaved perpendicular to the 
$c$-axis.  In our experiments, we studied a 25~$\mu$m thick $5 \times 
1.5$~mm$^2$ sample and a $1.5\times 2\times 4$~mm$^2$ sample (for $c$-axis response). 
Transmittance and reflectance measurements were 
performed using a Fourier transform spectrometer
in a frequency range from 10~to~45000~cm$^{-1}$ (1.2~meV to 5.6~eV).  
Temperature dependence from 4~to~300~K is achieved using liquid He in 
a continuous flow cryostat (sample in vacuum) with optical access 
windows.

\begin{figure}
\includegraphics[width=3.2in]{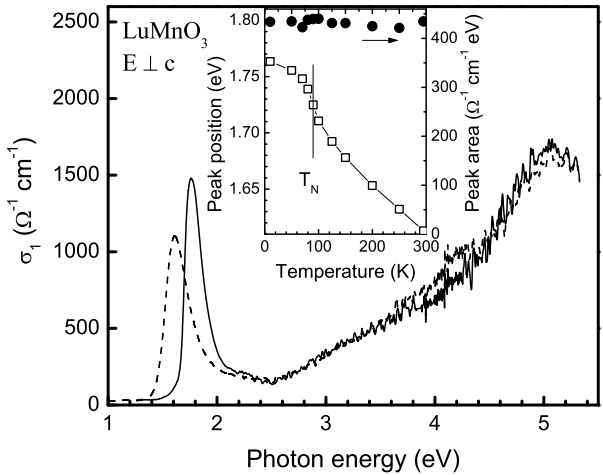}
\caption{\label{figvis} The electronic conductivity of LuMnO$_3$ at 
300~K (dashed line) and 10~K (solid line). Inset: temperature 
dependence of the peak energy (open squares) of the 1.7~eV feature and 
its spectral weight (solid circles).}
\end{figure}

The optical spectrum of LuMnO$_3$ is dominated by phonons in the 
28--100~meV photon energy range and electronic transitions starting 
at $\sim$1.1~eV. The material is transparent below 
28~meV and between 0.1--1.2~eV.

Figure \ref{figvis} shows the electronic part of the optical 
conductivity spectrum of LuMnO$_3$ calculated by Kramers-Kronig 
relations from the reflectance spectrum measured at 10 and 300~K. The 
lowest electronic excitation centered at $\sim$1.7~eV is seen to be 
strongly temperature dependent. A similar effect has been reported for 
hexa-YMnO$_3$ \cite{Kritayakirana1}. Whereas, in ortho-YMnO$_3$ this 
peak is not observed or is very weak \cite{Yi1}. From the Inset in 
Fig.~\ref{figvis} it is seen that the resonance energy decreases 
monotonically with temperature having an inflection point at 
$T_N\sim$~90~K while the integrated spectral intensity of this 
excitation is constant to within experimental accuracy of about 2~\%. 
This feature is sitting near the foot of a large spectral feature 
which has a weak but sharp onset at 1.1~eV and is maximum near 
$\sim$5~eV. The 1.1 to 5~eV conductivity band is independent of 
temperature to within our measurement accuracy.

We understand the electronic conductivity spectrum in terms of a broad 
band of charge transfer transitions from the hybridized oxygen 
$p$-levels to the Mn $d_{3z^2-r^2}$ levels centered at $\sim$5~eV and 
an on-site Mn $d$-$d$ transition centered at 1.7~eV.  The fact that 
this feature is observed in the hexagonal phase but not ortho-YMnO$_3$ 
is consistent with selection rules for the on-site Mn $d$-$d$ 
transitions in hexagonal and (near) cubic symmetry. Its interpretation 
as a charge transfer transition between Mn neighbors is ruled out 
because the spin dependence of the charge transfer matrix elements 
would lead to a strong temperature dependence of its oscillator 
strength which is not observed.

We have calculated the energies and spectral intensities of electronic
transitions in the MnO$_5$ complex in the framework of ligand-field
theory. The electronic states are taken to be the Mn $d$-orbitals
coupled to the oxygen $p_\sigma$ orbitals.  Taking the energy of the
$d_{x^2-y^2,xy}$ pair as zero, the energy of O $p$-orbitals is
$\Delta=-3$~eV, and the crystal field energy of the $d_{3z^2-r^2}$
orbital is $\Delta_{CF}=0.7$~eV.  For the symmetric (paraelectric)
structure, the hybridization is $t_1$$=$$t_2$$=$1.9~eV for apical and
$t_3$$=$$t_4$$=$1.7~eV for in-plane oxygens. The allowed optical 
transitions for the $E\perp c$ polarization of light are:
$d_{(x^2-y^2)',(xy)'} \rightarrow d_{(3z^2-r^2)'}$ at 1.58~eV and
from the two in-plane bonding O $p \rightarrow d_{(3z^2-r^2)'}$ at $\sim$6.3~eV,
where the prime denotes the corresponding $d$-states as 
hybridized with the oxygen orbitals.   
The calculated spectral  weights of both types of transitions are
a factor of four smaller than observed but their ratio is correct.
This discrepancy may be attributed to the neglect of the
Mn $p$- and $s$-orbitals which
can also couple to the O $p$ states in the hexagonal symmetry.  
In the case of $E||c$ polarization of incident light the optical matrix
elements for the $d_{(x^2-y^2)',(xy)'} \rightarrow d_{(3z^2-r^2)'}$ 
transitions are zero as we find experimentally.

What is the origin of the $\sim$0.15~eV shift of the 1.7~eV peak with 
temperature shown in Fig.~\ref{figvis}? The anomaly at $T_N$ indicates 
that at least part of the shift is associated with the magnetic phase 
transition. A spectral shift due to thermal expansion is also 
expected.  However, the observed shift is larger than is typically 
observed in interband features in solids ($\sim$~0.1~eV). We attribute 
the magnetic part of the shift to the effects of the exchange 
interactions between the Mn ions. Level shifts of the Mn $d$-levels 
due to superexchange between Mn neighbors leads to a lowering of the 
$d_{x^2-y^2,xy}$ levels in the antiferromagnetic state while the 
relatively isolated $d_{3z^2-r^2}$ orbital is little affected. We 
believe that  the shift in the resonance energy between 4~K and $T_N$ 
($\sim$0.05~eV) underestimates the exchange energy. This is partly 
because short range antiferromanetic correlations in this frustrated 
magnetic system are expected and are observed to persist to higher 
temperatures \cite{Sato1,Katsufuji1}.  Also, the absence of a shift in 
the 5~eV feature argues against such a large thermal shift in the 1.7~
eV feature since they share the same final state.  In addition, the 
extremely weak decrease in the oscillator strength of the 1.7~eV 
feature also argues against a strong thermal expansion effect since 
this optical transition, which is allowed only due to the 
hybridization with the O $p$ states, is more sensitive to the lattice 
constant than are the level shifts.  At present, however, we cannot 
separate the exchange effects from the effects of the thermal 
expansion on the optical transition energy.  This will require a 
detailed analysis of the thermal effects.  For completeness we note 
that magnetostriction can also produce level shifts due to the changes 
in the lattice constant.  However, these effects are much smaller than 
the effects of thermal expansion and can be safely neglected.

We can also estimate the exchange energy of the Mn spins from the 
Weiss temperature in the susceptibility.  From molecular field theory: 
$k_B\theta=z J_{NN} S(S+1)/3$ where $\theta$ is the Weiss temperature, 
$z$=6 is the number of nearest neighbors. Therefore the exchange 
energy for the manganese ion is $E_{ex}\cong 3 k_B\theta$.  In the 
literature  $\theta$ data for LuMnO$_3$ range between 
-519~K \cite{Tomuta1} and -887~K \cite{Katsufuji1}, so that $E_{ex}$ is 
in the range 140 to 240 meV, which is somewhat larger than our 
estimate from the optical shifts. As noted, however, this estimate 
ignored the additional shift expected above $T_N$ due to the effects 
of frustration.  Also, these two estimations represent different 
manifestations of the exchange interaction. The Weiss temperature 
represents the ground state exchange energy between the manganese 
moments.  While the optical shift represents the change in exchange 
energy of the Mn ion between the ground and excited state of the 
Mn$^{3+}$ ion. 
Therefore, the two estimates are in satisfactory agreement and our 
assignment of the magnetic shift of the 1.7~eV peak with temperature 
to the differences in exchange interaction in excited and ground 
optical states of a given Mn ion is reasonable.

\begin{figure}
\includegraphics[width=3.2in]{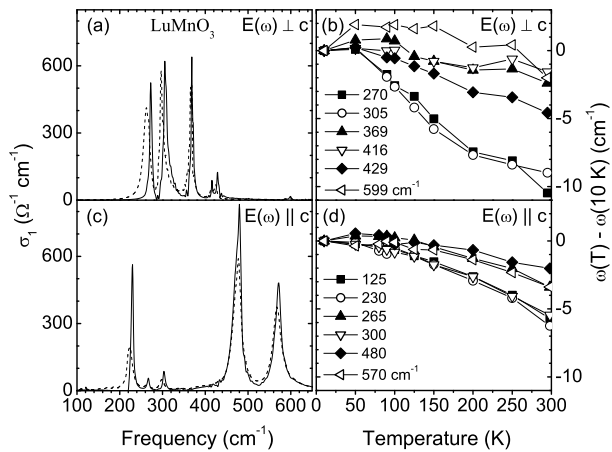}
\caption{\label{figphon} (a) and (c): Real part of the optical 
conductivity $\sigma_1$ of LuMnO$_3$ in the phonon spectral range for 
10~K (solid curves) and 300~K (dashed curves) for two polarizations of 
light; (b) and (d): temperature shifts of the phonon frequencies.}
\end{figure}

The phonon part of the conductivity spectrum of LuMnO$_3$ is shown in 
Fig.~\ref{figphon} (a) and (c). Mode assignments can be made by 
analogy with the assignment in YMnO$_3$ \cite{Iliev1}.  Panels (b) and 
(d) in Fig.~\ref{figphon} show the temperature dependence of the 
frequency shifts. The two low frequency vibrational modes in the $E 
\perp c$ polarization display the strongest absolute frequency shifts 
($\Delta\omega_0$), relative frequency shifts 
($\Delta\omega_0/\omega_0$), and inflection points at $T_N$. This 
observation suggests that these modes are coupled to the spin system. 
Measurements of the phonon spectrum in the $E || c$ polarization on a 
$ac$-plane sample show only a thermal shift of the phonons 
(Fig.~\ref{figphon}d).  The frequency shift of the spin-coupled phonon 
appears very similar to the temperature dependence of the 1.7~eV 
feature.  This suggests that both are related to the same, 
nearest-neighbor spin correlation function: $<\mathbf{S}_i\cdot 
\mathbf{S}_j>(T)$.

The phonon shifts are understood in terms of the phonon induced 
modulation of the exchange energy produced by the ion modal 
displacements. The change in exchange energy produces a corresponding 
change in the effective  restoring force for the phonon.  The 
resulting shift will be $\Delta\omega^2(T)=\Delta k/M$, where $\Delta 
k \sim <\mathbf{S}_i\cdot \mathbf{S}_j>$ is the exchange energy 
contribution to the force constant and $M$ is the reduced mass of the 
phonon mode. In addition to frequency shifts due to the exchange 
effects there will also be shifts due to the thermal expansion of the 
lattice. The exchange energy is dominated by superexchange between 
nearest neighbor Mn ions separated by oxygens.  However, due to the 
complexity of the hybridized Mn $d$-states there are both 
ferromagnetic and antiferromagnetic contributions that differ for in- 
and out-of-plane neighbors.

\begin{figure}
\includegraphics[width=3.2in]{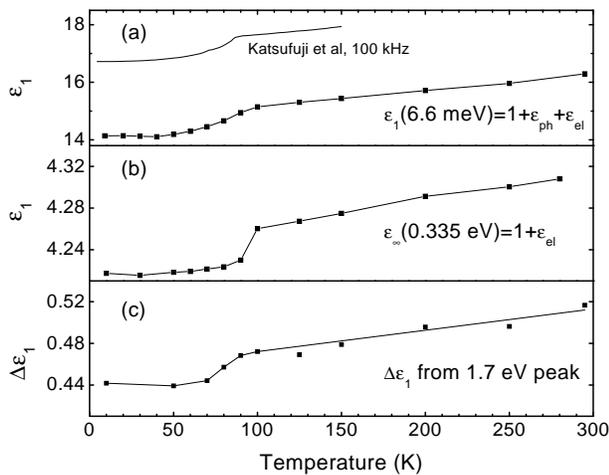}
\caption{\label{figeps} Temperature dependence of the dielectric 
constant $\epsilon_1$ of LuMnO$_3$. Top (a) and middle (b) panels 
represent $\epsilon_1$ measured from the etalon interference effect. 
Panel (c) shows the contribution of the 1.7~eV peak to $\epsilon_1$ 
determined from the optical sum rule. The lines are guides to the eye.}
\end{figure}

The anomaly in temperature dependence of the static in-plane 
dielectric constant below the N$\grave{e}$el temperature is considered 
to be one of the manifestations of coupling between magnetic and 
ferroelectric order parameters in the ferroelectromagnets 
\cite{Smolenskii1}. On the other hand, $\epsilon_0$ is determined by 
all of the oscillators present in the optical response of the system. 
We have examined the contributions to the quasi-static dielectric 
constant of LuMnO$_3$.  Three groups of oscillators contribute to 
$\epsilon_0$: ferroelectric domains, phonons, and electronic 
transitions. The contribution from the antiferromagnetic resonance, 
centered at 50~cm$^{-1}$ at 10~K, is negligible. The contributions 
from ferroelectric domains falls off with frequency becoming 
negligible at the MHz frequencies of the quasi static measurements 
and, even more so, at the far-infrared frequencies of our 
measurements.  Figure \ref{figeps} shows the temperature dependence of 
the real part of the dielectric constant of LuMnO$_3$. The top curve 
in panel (a) reproduces quasi-static data of Katsufuji {\it et al.} 
\cite{Katsufuji1}. The bottom curve in panel (a) was calculated from 
the measured frequency shift of the first interference maximum in the 
transmittance spectrum centered at 53~cm$^{-1}$. The value of 
$\epsilon_1$ at this frequency is determined by all optical phonons 
and all electronic oscillators. Panel (b) contains $\epsilon_1$ 
determined from the interference fringes in the mid-infrared 
transparency region midway between the phonon and electronic 
absorption bands. The value of $\epsilon_1 \equiv \epsilon_\infty$ at 
this frequency is due to the electronic transitions. Panel (c) shows 
an estimation of the input to the dielectric constant from the 1.7~eV 
electronic peak using $\Delta\epsilon_1=2/\pi 
\int_{1~eV}^{2.5~eV}d\omega\, \epsilon_2(\omega)/\omega$.  From the 
data of Fig.~\ref{figeps}, we conclude that practically all of 
$\Delta\epsilon$ below $T_N$ comes from phonon hardening.  Comparing 
panels (a) and (b), only $\sim$5~\% of this change is due to the shift 
of the 1.7~eV electronic peak. We can also conclude from panels (b) 
and (c) that the temperature dependence of $\epsilon_\infty$ is almost 
entirely due to the 1.7~eV feature.

The absence of a magnetic anomaly in phonons in the $E || c$ 
polarization is consistent with the observed anisotropy in the static 
dielectric constant \cite{Katsufuji1}. Toward developing an 
understanding of this anisotropy of the spin lattice coupling we note 
that there are important distinctions between the $E || c$ and $E 
\perp c$ polarizations in the exchange modulation for this layered 
system. In the case of $E || c$ polarization the relative displacement 
of the in-plane Mn and O ions is transverse to their bond and bond 
stretching occurs only with the apical oxygen, primarily affecting the 
Mn $z^2$ orbital which is empty.  Whereas for $E$ in the $ab$-plane 
both transverse and bond stretching Mn-O displacements are induced and 
the occupied in-plane orbitals are involved. This observation suggests 
that the in-plane bond stretching displacements dominate the 
spin-phonon interaction effects. The role of dynamic effective charge 
transfer for the bond stretching modes in transition metal compounds 
for ferroelectricity and the electron-phonon interaction has been 
recently discussed (see, for example, \cite{Resta1}). We suggest that 
the observed anisotropy of the spin-phonon coupling and $\epsilon_0$ 
might be a consequence of dynamic charge redistribution for in-plane 
vibrations of Mn ions, which modulate the in-plane partially covalent 
bonds, and the absence of the effective in-plane charge redistribution 
for the out-of-plane modes.  

More generally the question of a coupling between the two order 
parameters of this material is interesting.  Within the Landau theory 
of phase transitions there are symmetry allowed terms in the Landau 
free energy describing the coupling between the magnetic and 
ferroelectric order \cite{Smolenskii1}. The experimental data on the 
quasi-static dielectric constant implies that this term is of the form 
$\delta F \sim L^2P^2(E_x^2+E_y^2)$, where $P$ and $L$ are, 
respectively, the ferroelectric and antiferromagnetic order 
parameters. Establishing this coupling and relating it to the 
microscopic physics is a key issue in the study of this class of 
materials.

In conclusion, we have observed a strong coupling of the 
antiferromagnetism in LuMnO$_3$ to a sharp low energy interband 
transition and to the infrared phonon spectrum. The optical feature 
has a large blue shift associated with the antiferromagnetism which is 
caused by the effects of the exchange interaction on the on-site Mn 
$d$-$d$ transition. A similar anomaly in the temperature dependence 
of the phonon frequencies is attributed to effects of spin-phonon 
coupling. These results demonstrate that optical spectroscopy is a 
powerful tool in the study of exchange interaction in the strongly 
frustrated magnetic system of the hexagonal manganites.

\begin{acknowledgments}
This work supported in part by NSF-MRSEC DMR 0080008. Fruitful 
discussions with P.~B.~Allen and D.~B.~Romero are gratefully 
acknowledged. The work of J.~S.~Ahn is supported by the Korean Science 
and Engineering Foundation(KOSEF) through the Center for Strongly 
Correlated Materials Research (CSCMR) at Seoul National University.
\end{acknowledgments}

\end{document}